\documentstyle[12pt,equation]{article}
\begin{document}
\newcommand{\sigjet}{\mbox{$\sigma_{\gamma p}^{\rm jet}$}}
\newcommand{\sigtot}{\mbox{$\sigma_{\gamma p}^{\rm tot}$}}
\newcommand{\gamgam}{\mbox{$\gamma \gamma$}}
\newcommand{\gamp}{\mbox{$\gamma p$}}
\newcommand{\gamqq}{\mbox{$\gamma q \bar{q}$}}
\newcommand{\ptmin}{\mbox{$p_{T, \rm{min}}$}}
\newcommand{\ptch}{\mbox{$\langle p_{T, \rm{ch}} \rangle$}}
\newcommand{\nch}{\mbox{$\langle n_{\rm ch} \rangle$}}
\newcommand{\aem}{\mbox{$\alpha_{em}$}}
\newcommand{\phad}{\mbox{$P_{\rm had}$}}
\newcommand{\be}{\begin{equation}}
\newcommand{\ee}{\end{equation}}
\newcommand{\een}{\end{subequations}}
\newcommand{\ben}{\begin{subequations}}
\newcommand{\beq}{\begin{eqalignno}}
\newcommand{\eeq}{\end{eqalignno}}
\renewcommand{\thefootnote}{\fnsymbol{footnote} }
\noindent
\begin{flushright}
MAD/PH/841\\
June 1994
\end{flushright}
\vspace{1.5cm}
\pagestyle{empty}
\begin{center}
{\Large \bf Minijets in \gamp\ and \gamgam\ collisions\footnote{Invited talk
presented at the meeting on {\it Two--Photon Physics at LEP and HERA}, Lund,
Sweden, May 1994}} \\
\vspace{5mm}
Manuel Drees\footnote{Heisenberg Fellow}\\
{\em Physics Department, University of Wisconsin, Madison, WI 53706, USA}
\end{center}

\begin{abstract}
I discuss the minijet contribution to total photoproduction and
photon--photon cross sections. While minijets with $p_T$ around 2 GeV have
recently been observed directly in \gamgam\ experiments, the total \gamp\
cross section measured at HERA is in excellent agreement with predictions
based on purely soft physics. Due to the large number of free parameters,
predictions for the minijet contribution to total cross sections can
be brought into agreement with these seemingly paradoxical observations.
However, the currently used eikonalization procedure may not be applicable at
all to a large part of the minijet contribution, making it very difficult to
draw definite conclusions at present.
\end{abstract}

\clearpage
\noindent
\setcounter{footnote}{0}
\pagestyle{plain}
\setcounter{page}{1}
\section*{1) Introduction}
The idea that ``minijets", i.e. partonic jets with $p_T \sim 1-3$ GeV, drive
the observed increase of total hadronic cross sections with energy is now more
than 20 years old \cite{1}. While eikonalized minijet calculations can indeed
describe this increase \cite{2} it has to be admitted that a simple
power--law formula, based on ``old--fashioned" Pomeron physics, works just as
well \cite{3}. It had therefore been hoped that measurements of the total
photoproduction cross section at HERA ($\sqrt{s_{\gamma p}} \simeq 200$ GeV)
would serve to distinguish unambiguously between these two approaches.

This has unfortunately not happened. While the measurements by the H1 and ZEUS
collaborations \cite{4} fell just where estimates based on Pomeron exchange
predicted \cite{3} them to lie, proponents of the minijet idea where quick to
point out \cite{5,6} that these results are not at all inconsistent with a
large minijet contribution. Such ``post--dictions" might have been deemed to
be of dubious value if it were not for the fact that recently the TOPAZ and
AMY collaborations unambiguously observed \cite{7} minijets in \gamgam\
collisions at TRISTAN; see fig.1. These data, as well as other data on
multi--hadron production in \gamgam\ scattering \cite{8}, imply that partonic
collisions with transverse momenta in the GeV range occur at the rate
predicted by perturbative QCD.


It has long been known \cite{9} that the total {\em inclusive} cross section
for the photoproduction of such minijets reaches the level of the total \gamp\
cross section just about at HERA energies. This raises the challenging
question why this large inclusive cross section seems to make such a small
contribution to the total cross section, leading to the observed quite modest
increase of $\sigma^{\rm tot}_{\gamma p}$ between 20 and 200 GeV. In order to
tackle this problem, in sec. 2 I first describe current calculations of the
minijet contribution to total cross sections, which are based on
eikonalization. It quickly becomes apparent that these calculations contain
sufficiently many free parameters or even unknown functions to accomodate just
about any conceivable measurement of total cross sections, provided only they
rise with energy. What is worse, these calculations might still be too
simplistic; at least some of the underlying assumptions seem quite suspect to
me, as explained in sec. 3. Finally, sec. 4 contains a brief summary and
conclusions.

\section*{2) Eikonalized minijet cross sections}

The calculation of inclusive jet cross sections is a straightforward
application of perturbative QCD:
\be \label{e1}
\sigjet(p_T > p_{T, \rm{min}}) = \int_{x_{\rm min}}^1 d x_1
\int_{x_{\rm min}/x_1}^1 d x_2 f_{i|\gamma}(x_1) f_{j|p}(x_2)
\int_{p_{T, \rm{min}}}^{\sqrt{\hat{s}}/2} d p_T \frac {d \sigma_{ij}}
{d p_T}, \ee
with $x_{\rm min} = 4 p_{T,\rm{min}}^2/s$ and $\hat{s} = x_1 x_2 s$. Here
$f_{i|\gamma}$ and $f_{j|p}$ are distribution functions of partons $i$ and $j$
in the photon and proton, respectively, and $d \sigma_{ij}$ is the hard
scattering cross section of these partons. The cross section (\ref{e1}) grows
very quickly with energy. This can most easily be seen by assuming
$f_{i|\gamma}, \ f_{j|p} \propto x^{-(J+1)}$ at small $x$; using different
powers for the parton densities in the photon and proton does not change the
result qualitatively \cite{10}. Since $d \sigma_{ij} / d p_T \propto p_T^{-3}$
one has for $p_T^2 \ll s$:
\be \label{e2}
\sigjet (p_T > p_{T, \rm{min}}) \propto \frac{1}{J} \left( \frac {s} {4 p_T^2}
\right)^J \log \frac {s} {4 p_T^2}.
\ee
The power $J$ is expected to lie roughly in the range $0.2 \leq J \leq 0.5$.
The r.h.s. of eq.(\ref{e2}) therefore grows much faster with energy than total
cross sections do; experimentally, $\sigma^{\rm tot} \propto s^{0.08}$ for
both $\bar{p} p$ and \gamp\ collisions \cite{3}.

Recall, however, that eq.(\ref{e1}) describes an {\em inclusive} cross
section. This differs from the minijet contribution to the {\rm total} cross
section by the average number of jet pairs (or partonic collisions) per
hadronic collision:
\be \label{e3}
\sigma_{\gamma p}^{\rm{jet, tot}} = \sigjet / \langle n_{\rm jet} \rangle.
\ee
In the usual eikonalization scheme \cite{2,5,6,10} the possibility of
producing more than one jet pair in a \gamp\ collision is included by assuming
that several parton--parton collisions occur {\em independently} of each
other; the number of partonic collisions per event then obeyes a Poisson
distribution. In order to estimate the value of $\langle n_{\rm jet} \rangle$
one also has to know the transverse overlap of the parton densities. In the
usual treatment one makes the second crucial assumption that the dependence
of the parton densities on Bjorken$-x$ and on the impact parameter $b$
factorizes.

Under theses assumptions the total \gamp\ cross section can be computed from
\cite{10}:
\be \label{e4}
\sigma_{\gamma p}^{\rm tot} = \phad \int d^2 b \left[ 1 - \exp \left(
- \frac {\chi_{\rm soft} + \sigma_{\gamma p}^{\rm jet} (s) } {P_{\rm had}}
A(b) \right) \right].
\ee
Here, \sigjet\ is given by eq.(\ref{e1}). $\chi_{\rm soft}$ is the ``soft"
(non--perturbative) contribution to the eikonal; it is mostly determined by
low--energy data, but is has recently been argued \cite{6} that it might
show nontrivial $s-$dependence even at high energies. The function $A(b)$
describes the transverse overlap of the parton densities; it is normalized
such that $\int d^2 b A(b) = 1$.

Finally, the parameter \phad\ appearing in eq.(\ref{e4}) is supposed to
describe the probability for a photon to go into a hadronic state. This is
clearly ${\cal O}(\aem)$, but the exact value is not known. The necessity to
introduce such a parameter has first been pointed out in ref.\cite{10}. A
very intuitive argument has been given in ref.\cite{11}; it is based on the
expansion of eq.(\ref{e4}) for small \sigjet. The $n.-$th term in this
expansion describes the cross section for the simultaneous production of $n$
jet pairs. This gives:
\be \label{e5}
\sigma(n \ {\rm jet \ pairs}) \propto \phad \left( \frac
{\sigma_{\gamma p}^{\rm jet}} {P_{\rm had}} \right)^n.
\ee
Notice that \sigjet\ in eq.(\ref{e1}) is ${\cal O}(\aem)$, since the
$f_{i|\gamma}$ are ${\cal O}(\aem)$. If $\phad=1$ the cross section for
producing $n$ jet pairs would therefore be ${\cal O}(\alpha^n_{em})$. This is
counter--intuitive; once the transition into a hadronic state has been made,
no further electromagnetic interactions are needed to produce additional
jet pairs. On the other hand, if $\phad \sim {\cal O}(\aem)$, eq.(\ref{e5})
gives $\sigma (n \ {\rm jet \ pairs}) \sim {\cal O}(\aem)$, as expected.

Clearly a great number of a priori unknown parameters and functions has to be
fixed before eq.(\ref{e4}) can be evaluated. To begin with, the jet cross
section \sigjet\ depends on the parton densities, especially at small $x$
and $Q^2 \sim p^2_{T, \rm{min}}$. At least in principle these densities can
be measured in processes that can be described by purely perturbative QCD. In
contrast, \ptmin\  is clearly not computable from perturbation theory alone;
in fact, by introducing this parameter one hopes to describe the intricacies
of confinement by a single parameter. A very similar cut--off parameter has
been introduced in analyses \cite{7,8} of multi--hadron production in \gamgam\
reactions. Unfortunately different groups find different preferred values of
\ptmin\ even if they use the same set of structure functions. E.g., for the
old DG parametrization \cite{12}, DELPHI finds a value as low as 1.45 GeV,
while ALPEH data seem to favor \ptmin\ around 2.5 GeV; results from AMY and
TOPAZ lie in between. Fig.2 shows that such a variation of \ptmin\ changes
predictions for the minijet contribution to \sigtot\ by at least a factor of 2
even at very high energies.


As discussed above, \phad\ has to be ${\cal O}(\aem)$ for eq.(\ref{e4}) to
make sense at all; however, the exact value is unknown. The original estimate
of ref.\cite{10} was $\phad = 4 \pi \aem / f^2_{\rho} \simeq 1/300$, but
later a value of $1/170$ has been suggested \cite{11} based on parton model
considerations. Fig. 3 shows that a factor--of--two uncertainty in \phad\ also
leads to a substantial uncertainty in the prediction for \sigtot.


$\chi_{\rm soft}$ is usually written in the form $ A + B / \sqrt{s}$, making it
independent of $s$ at large energies. However, as already mentioned above,
it has recently been suggested \cite{6} that $\chi_{\rm soft}$ might also
grow slowly with energy. This will obviously affect predictions for \sigtot.

Finally, the function $A(b)$ needs to be specified. In all minijet
calculations of \sigtot\ that I am aware of $A(b)$ has been assumed to be the
Fourier transform of the product of the electromagnetic form factors of the
proton and of the pion. The use of electromagnetic form factors to estimate
the transverse distribution of partons in the proton is certainly not
unreasonable, although it does not allow for the occurence of ``hot spots".
Using $\pi$ form factors as an estimate of the transverse parton distribution
in the photon is quite a different matter, though. This approach is based on
the VDM assumption that a photon is ``basically" a vector meson (or a
superposition of $\rho, \ \omega, \ \phi$ and higher states), and the
additional assumption that the $\rho$ is ``basically" like a $\pi$. To begin
with, the $\rho$ meson is really not much like a pion at all, being about 5
times heavier; indeed, since the $\rho$ is a resonance, it can even be
described as ``consisting" of two pions! More seriously, we know
experimentally that the $x-$dependence of the quark distribution functions in
the photon does {\em not} look like that of the pion at $Q^2 \sim
p^2_{T,\rm{min}} \sim$ (a few) GeV$^2$. In my view there is therefore no
reason to assume that the $b$ dependence is similar for the photon and the
pion.

The difference in the $x-$dependence of photonic and pionic parton densities
is largely due to the hard \gamqq\ coupling. The existence of this pointlike
vertex suggests to estimate the transverse distribution of partons in a
photon as a Fourier transform of the hard intrinsic $k_T$ distribution of the
quarks produced in this vertex:
\beq \label{e6}
q^{\gamma}(b) &\propto \int d^2 k_T \frac {\exp(-i \vec{k_T} \vec{b})}
{k_T^2 + k_0^2} \nonumber \\
&\propto \int d k_T \frac {k_T}{k_T^2 + k_0^2} J_0(k_T b),
\eeq
where $k_0$ is an IR regulator and $J_0$ is a Bessel function. The
distribution (\ref{e6}) is peaked at $b=0$; more importantly, its width is
given by the inverse of the hard momentum scale in the problem, as opposed to
the (rather large) radius of the pion. In other words, one expects this
(hard) contribution to the parton densities in the photon to be more strongly
peaked in transverse direction than in case of the pion. A narrower
distribution means a larger $A(0)$, which increases eikonalization effects,
i.e. reduces the predicted minijet contribution to \sigtot. A similar
connection between $A(b)$ and the intrinsic $k_T$ of the partons in the
photon has been incorporated in the latest refinement \cite{13} of the
Schuler--Sj\"ostrand model \cite{14} of photonic interactions; however, this
model does not attempt to predict the total \gamp\ cross section (although it
does predict the relative size of various contributions to that cross
section).


\section*{3) Is eikonalization applicable at all?}

The discussion at the end of the previous section raises doubts whether
eikonalization should be used at all for those resolved photon contribution
that come from the hard (perturbative) part of the photon structure functions.
Recall that one of the fundamental assumptions in the derivation of
eq.(\ref{e4}) was that multiple parton--parton reactions can occur
{\em independently} in one \gamp\ scattering event. On the other hand, the
entire perturbative part of the parton densities in the photon can by
definition be traced back to the \gamqq\ vertex. Given that all these partons
manifestly originate from a common source it seems unlikely that they can be
treated as being statistically independent. This is illustrated in fig. 4 for
the case of two jet pairs. The sum of diagrams of the type shown on the left
is supposed to be equal to the $\gamma \rightarrow$ hadrons transition
probability multiplied with the square of the diagram to the right. Notice
that it is assumed here that the first step, the $\gamma \rightarrow$
hadrons transition, can simply be described by a constant; in other words, the
parton densities describing this ``hadronic state" are assumed to have the
same $x$ and $Q^2$ dependence as the usual photonic parton densities, up to a
constant factor. This is clearly a crude approximation at best.

Given that the applicability of eikonalization to a large part of resolved
photon contributions is doubtful, it seems to be a good idea to look for
experimentally measureable quantities that are sensitive to the existence of
minijets but are {\em not} sensitive to eikonalization. Such quantities should
therefore depend on the perturbatively calculable {\em inclusive} minijet
cross section (\ref{e1}), rather than on its contribution to the total
cross section.

Some time ago I suggested \cite{14a} that the product $\sigma^{\rm tot, inel}
\cdot \nch \cdot \ptch$ might be a good candidate for such a quantity, where
$\sigma^{\rm tot, inel}$ is the total inelastic cross section, \ptch\ the
average $p_T$ of charged particles and \nch\ the average charged particle
multiplicity. The energy dependence of this quantity as measured in $\bar{p} p$
collisions is depicted in fig. 5; at least over the range shown here it seems
to be described quite well by a simple linear function. This rapid increase is
quite consistent with the rapdi rise (\ref{e2}) of the inclusive minijet cross
section. Notice that each of the three factors is predicted by minijet models
to increase with energy. Indeed, one of the strengths of the minijet model is
that it allows to estimate such quantities at all; its usefulness therefore
goes well beyond the prediction of total cross sections. The prediction for
each factor by itself depends on the eikonalization scheme, but the product
should not depend on this: It should not be important whether two pairs of
minijets are distributed over two \gamp\ events (giving large $\sigma^{\rm
tot, inel}$ but small $\ptch \cdot \nch$) or are concentrated in one event
(giving small $\sigma^{\rm tot, inel}$ but large $\ptch \cdot \nch$).
Unfortunately, this quantity is sensitive to fragmentation effects \cite{15},
since the scalar sum of the $p_T$ of all (charged) particles does not add up
to the $p_T$ of the parton producing a minijet.


Nevertheless this (or a similar) quantity should be useful for determining the
only nonperturbative parameter entering the calculation (\ref{e1}) of
inclusive minijet rates, i.e. \ptmin; here I assume that the relevant parton
densities will be determined from other reactions (DIS, $c \bar{c}$ and
$J/\psi$ production, \dots). The {\em inclusive} minijet cross section will
then be known, and we can try to figure out how these minijets are
distributed over \gamp\ events by studying details of these events, as done
in ref.\cite{16} for $\bar{p} p$ collisions. One possible problem of this
approach is that the eikonalization ansatz (\ref{e4}) contains so many
free parameters that it might be able to describe a large amount of data even
if it is intrinsically flawed. Still, it seems clear to me that at present
it is hopeless to try and make predictions for \sigtot\ based on minijet
models unless one uses either additional data or additional theoretical
assumptions \cite{5} as input; either way one has to go beyond the realm of
perturbative QCD. Of course, the same remarks that I made here for \gamp\
scattering also apply for \gamgam\ reactions.

\section*{4) Summary and conclusions}
Minijets exist (see fig. 1), but at present we are not able to compute their
contribution to the total \gamp\ cross section reliably. As shown in sec. 2
the usual eikonalization prescription contains many unknown parameters. Even
worse, in sec. 3 I have presented arguments casting doubt on the validity of
this formalism for contributions coming from the perturbative part of photon
structure functions. Unfortunatly at present I cannot offer any alternative
scheme to compute total cross sections from inclusive jet cross sections. It
seems clear, though, that we have to use much more experimental information
to determine, first, the {\em inclusive} minijet cross section at high
energies, and in a next step, to figure out how these minijets are distributed
over \gamp\ or \gamgam\ events. Only time will tell whether such a program
will eventually force us to abandon conventional eikonalization schemes for
\gamp\ and \gamgam\ reactions.

\noindent
\subsection*{Acknowledgements}
I thank the organizers for inviting me to this very well planned and executed
meeting. This work was supported in part by the U.S. Department of Energy
under contract No. DE-AC02-76ER00881, by the Wisconsin Research Committee with
funds granted by the Wisconsin Alumni Research Foundation, as well as by a
grant from the Deutsche Forschungsgemeinschaft under the Heisenberg program.

\clearpage
\section*{Figure Captions}

\renewcommand{\labelenumi}{Fig.\arabic{enumi}}
\begin{enumerate}

\item 
Example of reconstructed minijets observed by the TOPAZ collaboration [7].

\vspace*{5mm}
\item  
Dependence of the minijet contribution to \sigtot\ on
\ptmin. The curves are for $\ptmin=2,3,4,5$ GeV. From ref.[13].

\vspace*{5mm}
\item  
Dependence of the minijet prediction for \sigtot\ on \phad; from ref.[11].

\vspace*{5mm}
\item  
In the usual eikonalization scheme, the diagram contributing to the
simultaneous production of two minijet pairs (left) is supposed to be
described by the square of the diagram shown on the right, multiplied with a
constant $\gamma \rightarrow$ hadrons transition probability.

\vspace*{5mm}
\item  
Energy dependence of $\sigma^{\rm tot, inel} \cdot \nch \cdot \ptch$
as measured at $\bar{p} p$ colliders; from ref.[16].

\end{enumerate}

\clearpage
\begin{thebibliography}{99}

\bibitem{1}
D. Cline, F. Halzen and J. Luthe, Phys. Rev. Lett. {\bf 31}, 491 (1973).

\bibitem{2}
For example, M. Block et al., Phys. Rev. {\bf D41}, 978 (1990).

\bibitem{3}
A. Donnachie and P.V. Landshoff, Phys. Lett. {\bf B285}, 172 (1992).

\bibitem{4}
H1 collab., T. Ahmed et al., Phys. Lett. {\bf B299}, 374 (1993);
ZEUS collab., M. Derrick et al., Phys. Lett. {\bf B293}, 465 (1992) and
DESY report 94--032.

\bibitem{5}
R.S. Fletcher, T.K. Gaisser and F. Halzen, Phys. Lett. {\bf B298}, 442
(1993).

\bibitem{6}
J.R. Forshaw and J.K. Storrow, Phys. Lett. {\bf B321}, 151 (1994).

\bibitem{7}
TOPAZ collab., H. Hayashii et al., Phys. Lett. {\bf B314}, 149 (1993);
AMY collab., B.J. Kim et al., Phys. Lett. {\bf B325}, 248 (1994).

\bibitem{8}
AMY collab., R. Tanaka et al., Phys. Lett. {\bf B277}, 215 (1992);
ALEPH collab., D. Buskulic et al., Phys. Lett. {\bf B313}, 509 (1993);
DELPHI collab., P. Abreu et al., CERN report PPE--94--04; see also the
talks by A. Finch and F. Kapusta, these proceedings.

\bibitem{9}
M. Drees and F. Halzen, Phys. Rev. Lett. {\bf 61}, 275 (1988).

\bibitem{10}
J.C. Collins and G.A. Ladinsky, Phys. Rev. {\bf D43}, 2847 (1991).

\bibitem{11}
R.S. Fletcher, T.K. Gaisser and F. Halzen, Phys. Rev. {\bf D45}, 377 (1992);
erratum, {\bf D45}, 3279 (1992).

\bibitem{12}
M. Drees and K. Grassie, Z. Phys. {\bf C28}, 451 (1985).

\bibitem{12a}
J.R. Forshaw and J.K. Storrow, Phys. Rev. {\bf D46}, 4955 (1992).

\bibitem{13}
G.A. Schuler, these proceedings.

\bibitem{14}
G.A. Schuler and and T. Sj\"ostrand, Nucl. Phys. {\bf B407}, 539 (1993).

\bibitem{14a}
M. Drees, Proceedings of the {\it International Workshop on Photon--Photon
Collisions}, San Diego, Calif., March 1992.

\bibitem{15}
R.S. Fletcher, private communication.

\bibitem{16}
T. Sj\"ostrand and M. van Zijl, Phys. Rev. {\bf D36}, 2019 (1987).

\end{thebibliography}
\end{document}